\documentclass[twocolumn,widetext,preprintnumbers,amsmath,amssymb]{revtex4}
\usepackage{graphicx}
\usepackage{dcolumn}
\usepackage{bm}
\usepackage{amsmath, amssymb}
\begin{document}
\title{Random planar graphs and the London street network}
\author{A. P. Masucci, D. Smith, A. Crooks, M. Batty
}                     
%
%
\affiliation{Centre for Advanced Spatial Analysis, University College London, 1-19 Torrington Place, London, WC1E 7HB, UK}
\date{\today}
%
\begin{abstract}
In this paper we analyse the street network of London both in its primary and dual representation. To understand its properties, we consider three idealised models based on a grid, a static random planar graph and a growing random planar graph. Comparing the models and the street network, we find that the streets of London form a self-organising system whose growth is characterised by a strict interaction between the metrical and informational space. In particular, a principle of least effort appears to create a balance between the physical and the mental effort required to navigate the city.
\end{abstract}
\pacs{89.75.-k, 89.75.Da, 89.65.Lm}
\maketitle
\section{Introduction}
\label{intro}

Urban growth has been widely analysed in the last century using ideas from social physics and urban economics \cite{20}. In fact cities, as natural phenomena, provide an iconic paradigm for the science of complexity, both with respect to their allometric scaling laws that relates them to the celebrated Zipf's law for population ranks \cite{21} and for the complexity of their transport patterns that have been analysed both in the context of fractal geometry \cite{22} and network theory \cite{8,16,2}.

Graph theory provides a natural environment to study urban growth as far back as 1736 , Euler applied graph theory to solve an urban problem, the well known K\"{o}nigsberg bridges problem \cite{12}, thus relating a metrical problem to a topological one.

A graph $\mathfrak G$ is a very simple object, i.e. an ensemble of $V$ vertices representing objects and $E$ edges representing the relations between the objects, $\mathfrak G=\{V,E\}$. With this level of abstraction, graphs have been applied in geographical studies in different ways, for instance to study the patterns of urban commuting \cite{10}, the spread of infectious diseases \cite{23} and networks of the retail system \cite{9}.

If we assume that the vertices of a graph are the street intersections in a city and the extremes of \emph{dead end roads} (or cul-de-sacs) and the edges the street fragments connecting the intersections, we obtain a so-called \emph{street network}. In particular we call this representation a \emph{primary representation} of the street network following the terminology in \cite{3}. Such a street network is a strange network when compared to other social or biological networks \cite{15} in the sense that it is embedded in the Euclidian space and the edges do not cross each other. In graph theory, such a network is called a \emph{planar graph} \cite{4}.

 The study of planar graphs has not received much attention in physics for two main reasons. The first is that the planarity criteria is not easy to overcome using the calculus. Therefore a lack of analytical results has discouraged analysts  in dealing with such graphs. The second is that planar graphs can appear trivial in both their topological and  geometrical properties. Regarding the first issue, we believe that since planar graphs represent a class of important phenomena, simulations can be used to quantify the basic properties of such graphs.  Regarding the second issue, we note that the current research in the field is limited to static planar graphs. In this paper, we introduce a new class of models of growing planar graphs that show more articulated properties than  their static counterparts.

Moreover in the study of street networks, there is considerable interest in the so-called \emph{dual representation}, that is the representation in which the streets are vertices and two vertices are connected whenever the streets they represent intersect\footnote{It is worth to notice that in graph theory the dual representation of a planar graph has a different meaning. In particular for a planar graph $\mathfrak{G}$, the dual graph $\mathfrak{F}$ is the graph in which the faces of $\mathfrak{G}$ are the vertices and two vertices  are connected whenever the faces they represent share the same boundary in $\mathfrak{G}$. Nevertheless in this paper we follow the definitions introduced in physics reviews.}. This representation describes the information content of the street network \cite{1}, in the sense that it represents the way a person navigates the city. To understand this concept, we need to refer to our personal experience when we  move from one place to another in the city. In such a case, we do not think of all the street segments we cross to go  from one point to another, but only the roads we move on, that are the vertices of the dual representation. Hence  to cross a large city (like London), we only need a small amount of information such as the street names (the vertices of the dual representation) which we need to cross the city.

This concept will become clearer later. For now it is important to mention that it has been observed that the distribution of the number of connections (the \emph{degree distribution}) of the vertices of the dual representation of street networks is often scale-free \cite{1}. This observation relates the phenomenology of urban growth to a wide range of scale-free phenomena through network theory representation and allows us to think of the growth of a city using an informational approach.

In this paper, we analyse the street network of London in its primary and dual representation. To contextualise the  results, we first introduce a grid model to simulate a maximal ordered city and then two stochastic models, one static and one a growth model, to simulate a maximal random city. In the primary representation, we construct measures in the topological  and metrical space, in the cycle space, and in the information space. In the dual representation, we generate measures in the topological and information space. Notably we find that the structure of London streets tends to be a compromise  between a growing random city and a grid-like city in the sense that it is self-organised in a way that minimises both  the physical and the informational effort required in navigating the city.

The importance of this research resides first of all in the quality of the analysed data (see appendix A for details),
then in the detailed analysis of static planar graphs, and lastly in  the introduction and analysis of growing random planar graphs

\subsection{The street network of London}

\begin{figure}[!ht]\center
                \includegraphics[width=0.5\textwidth]{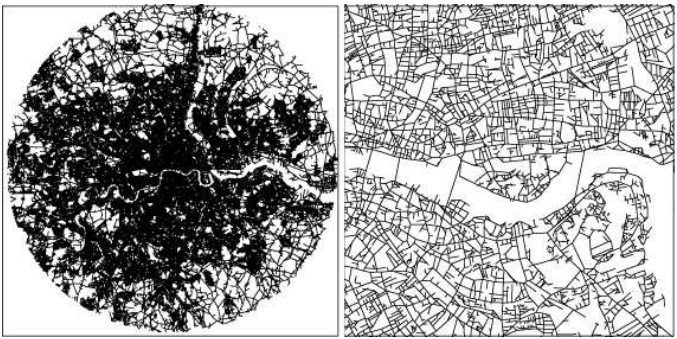}
 \caption{\label{i2} Left panel: the London street network considered in this research. Right panel: a localised view of the same network.}
    \end{figure}

London began in 43AD as a Roman settlement and has had comparatively uninterrupted urban growth every since making it the largest metropolis in Western Europe. To establish the borders of a city is still a controversial topic \cite{28} and hence, to build our network, we consider all the streets contained in a circle of radius 28.26 Km, centred on the centroid of the borough called \emph{the City of London}, where the first Roman settlement was located. This area contains some 95 percent of the population of the 33 boroughs that comprise the Greater London Authority which is also bounded by the M25 orbital road. In this way, we obtain a network with $V = 163878$ intersections, the vertices, and $E = 199931$ street segments, the edges (see the left panel of Fig.\ref{i2}). The London street network (hereafter LN) is a weighted network where the weights $w_{ij}$ of the edges connecting vertex $i$ to vertex $j$ are defined by the length $l$  of the street fragment they represent. A key measure for such a network is the degree $k_i$ of a vertex $i$ defined as the number of vertices vertex $i$ is connected to. The average degree for LN is $<k>=\frac{2E}{V}\simeq2.44$, a  very  small value, close to that of a tree, it is due to the massive presence of dead end roads as we can see from the right panel of Fig.\ref{i2}.

\begin{figure}[!ht]\center
                \includegraphics[width=0.5\textwidth]{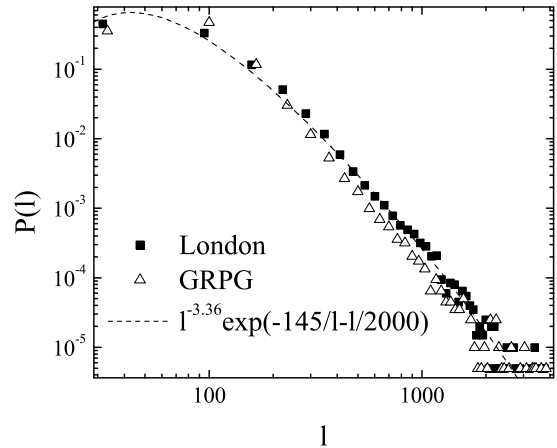}

 \caption{\label{i3} Measure of the length distribution $P(l)$ for the street network of London and for the GRPG (growing random planar graph).}
    \end{figure}

 An important measure that we will use in the next section is the density distribution of the length $l$ of the street segments, i.e. the weight distribution of LN, measured in meters. We show it in Fig.\ref{i3} and we find that it is well fitted by the function:
\begin{equation}\label{1}
f(l)\propto exp\left[-\frac{145}{l}-\frac{l}{2000}\right]l^{-3.36},
\end{equation}
where the average length for an edge is $95.73mt$. The properties of Eq.\ref{1} are scale-free for a long range of distance, and the long distance cut-off ensures that the variance of the distribution is finite.

\subsection{The Erd\"{o}s-R\'{e}nyi Random Planar Graph }

We first introduce a random model for a static planar graph. This is the only kind of random planar graph considered in literature as far as we know and we follow convention in calling it the Erd\"{o}s-R\'{e}nyi planar graph (hereafter  ERPG) in \cite{5}.

To build an ERPG  we start with a Poisson distribution of $N$ points in a plane and we choose  a distance $r$. To build the first segment, we randomly pick up two points of the distribution that have a distance less then $r$ and we connect them. Then we continue to randomly pick up pairs of points $P$ and $Q$ in the given points distribution that have a distance less then $r$. If the segment $\overline{PQ}$ does not intersect any other line of the graph, we add it to the graph. The process ends when we add the desired number of edges $E$ or when we arrive to the maximum allowed number of  edges $E\leq\frac{13}{7}V-o(V)$\cite{6}.

\begin{figure}[!ht]\center
                \includegraphics[width=0.5\textwidth]{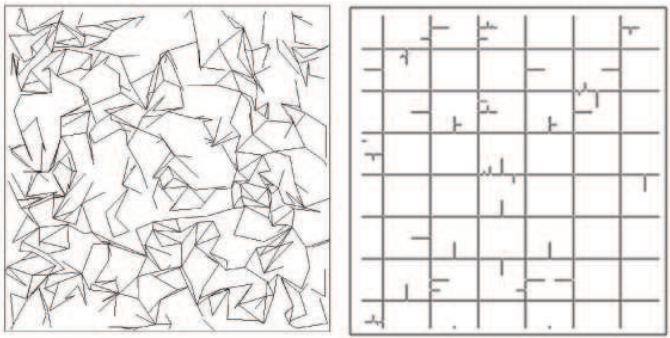}

 \caption{\label{i1} Left panel: a localised view of a realisation of the ERPG. Right panel:  a localised view of a realisation of the Grid Model with degree $<k>=2.44$. }
    \end{figure}

Here we generate a realisation of the ERPG model with the same characteristics as the LN, that is the same number of vertices and edges, and a distribution of points in a disc with the same radius as the LN. To obtain the same average length for the links, we choose $r = 300mt$. A localised view of this realisation is shown in the left panel of Fig.\ref{i1}, where we should note that this graph is not necessarily fully connected. In particular, the realisation we took as a study sample is made of 2072 disconnected components, the largest one composed of 146965 vertices.

 \subsection{The Growing Random Planar Graph }

The ERPG is a static model for a planar graph. Since cities are often growing systems that assume their shape over the centuries, we introduce a novel class of random planar graphs which we call growing random planar graphs (hereafter GRPG). We will show how the growth of this graph implies different emerging properties from the ERPG.

To build a  GRPG we start with a segment of length $\lambda$ embedded in the Euclidean plane. At each time step, we randomly pick up one of the vertices of the  graph. We draw from it a new segment of length $l$ according to an isotropic distance distribution $f(l,\Theta)=f(l)$, where $f$ is a probability density function. If the new segment does not intersect any of the existing segments, then we add it to our graph. This process creates a tree planar graph with average degree $<k>=\frac{2E}{N}=\frac{2(V-1)}{V}$. To obtain a planar graph that is not a tree and that has average degree $<k> >2$, every $n$ time steps we randomly pick  a  vertex $i$ from the existing graph. Next we consider the set of vertices in the graph that are within a radius $l_0$ from vertex $i$, where $l_0$ is randomly extracted from the distribution $f(l)$,  and  forms a segment with vertex $i$ that does not intersect  any other segment of the graph.  Then we randomly pick up a vertex $j$ from this set of vertices and we add the line $\overline{ij}$ to the graph. The process continues until we reach the desired number of edges or vertices.
The average degree of the vertices is then completely determined by $n$, $<k>=2+2/n$ and thus the GRPG properties are completely determined by the choice of $n$ and $f(l)$.

\begin{figure}[!ht]\center
                \includegraphics[width=0.5\textwidth]{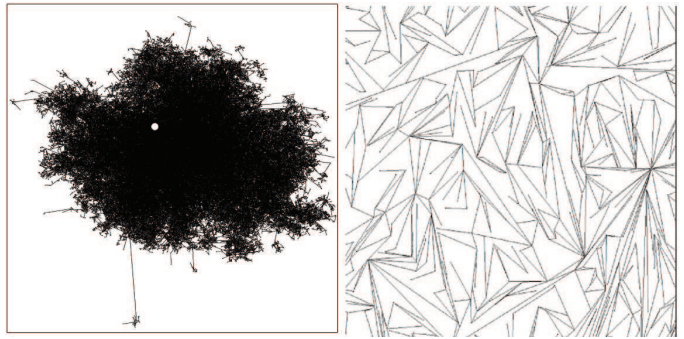}
 \caption{\label{i4} Left panel: the realisation of the GRPG  considered here, where the white dot is the origin of the growth of the model. Right panel: a localised view of the same network.}
 \end{figure}

Here we analyse a realisation of a GRPG with the same number of vertices and edges as the LN, $f(l)$ given by Eq.\ref{1} (see Fig.\ref{i3}) and $n=5$. We show this realisation in Fig.\ref{i4}, where the white dot shows where the first segment was located. We also notice how the power law distribution for the length of the edges allows long range connections, thus creating independent centres outside of the main cluster city which leads to an overall asymmetric form. Changing the distribution $f(l)$ it is possible to obtain different shapes of cities. Moreover in the GRPG there are no unconnected components as in the ERPG.

\subsection{The Grid }

The last model we introduce is that of a regular grid (GM hereafter) to which we randomly add dead end roads to obtain the same average degree of the LN. We introduce this graph to simulate a maximally ordered city.

We start with a square grid of $n$ horizontal lines and $n$ vertical lines. As in the previous networks, the vertices are defined by the intersections of the lines and the edges of length $l$ by the lines connecting two intersections. In this way, we create $n^2$ vertices with degree 4. To create the same average degree of the LN, for $m$ time-steps, we add a new line in the following way. We randomly pick up an edge from the network and from  its midpoint we draw a new line of length $l/2-\sigma$, perpendicular to the selected edge, where $\sigma=o(l)$. This process creates 2 new lines and two new vertices of degree 1 and 3 at each time-step. The resulting network has
 \begin{equation}\label{3}
 V=n^2+2m
\end{equation}
 vertices and it is easy to show that the average degree of the network is given by the following relation:
\begin{equation} \label{4}
<k>=4\frac{(n^2+m)}{n^2+2m}.
\end{equation}
Hence to find the correct values of $n$ and $m$ in building our grid model, it is sufficient to solve the system of equations \ref{3} and \ref{4} with the values of $V$ and $<k>$ taken by the LN, and we find $n^2=36053.2$ and $m=63912.4$. Considering that we need integer numbers, we run a simulation with $n=190$ and $m=63912$ and this gives us the same average degree as LN. In the right panel of Fig.\ref{i1} we show a localised realisation of such a graph.

 \section{Comparison between the London street network and the different models in their primary representation}

 In this section we  compare the properties of the LN, the ERPG, the GRPG and the GM introduced in the last section.  This section is divided in three subsections where we study the topological and geometrical properties, the measures in the cycle space, and the centrality measures which are all analysed separately. Many of the measures regarding the ERPG and the GM are trivial and are not considered.

\subsection{Topological and geometrical properties}

In a planar graph, topological and geometrical properties are very much interrelated. We begin by considering a geometrical feature, the spatial density of intersections $\rho$. The density of the intersections, or vertices, is an emergent property of the complex organisation of a growing planar graph. In the case of the ERPG it is Poisson, while in the case of the GM it is a uniform distribution.

\begin{figure}[!ht]\center
                \includegraphics[width=0.5\textwidth]{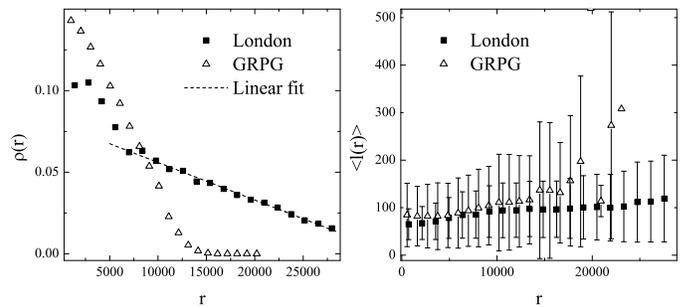}
 \caption{\label{i6} Left panel: the radial density $\rho(r)$ of intersections  for the LN and  the GRPG. The tail of the measure for London  is well fitted by a linear function (Adj. $R^2=0.99029$). Right panel: measures of the average edge length $l(r)$ versus the distance from the centre for the LN and for the GRPG. }
    \end{figure}

In the left panel of Fig.\ref{i6}, we show the measure of the radial density $\rho(r)$ of the intersections in LN compared to the one measured in the GRPG. In the case of LN, we see that $\rho(r)$ has a density plateau up to a radius of approximately 3.5Km, then the density drops fast until a radius of around 7Km from the centre is reached. After that, the behaviour changes abruptly and $\rho(r)$ decays linearly toward the periphery. In the case of the GRPG, we can see that the growth of the graph produces a density distribution that is a smooth bell shaped decaying function of the distance. The plateau that is in LN is missing and the function decays rapidly to a radius around of 15Km producing a random city that has an extension that is a half of its real counterpart. The linear decay of the density function for LN is related to the city's historical suburban growth and can be related to the phenomena we call \emph{urban sprawl} \cite{35}.

\begin{figure}[!ht]\center
                \includegraphics[width=0.5\textwidth]{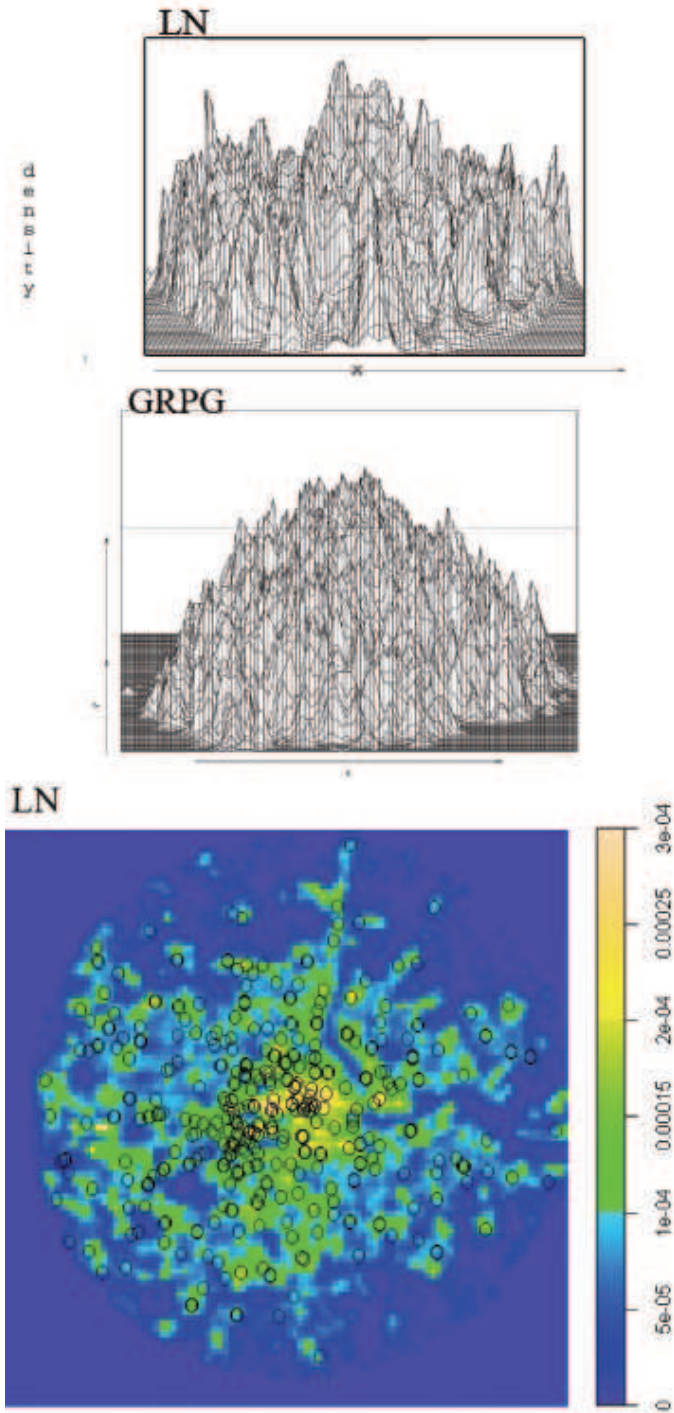}
 \caption{\label{i5} Upper panel: the intersection density profile for the LN. Central panel: the same measure for the GRPG. Bottom panel: the density contours for the LN. The black circles show the current (2006) position of the main town centres. }
    \end{figure}

  This behaviour can be better understood if we look at Fig.\ref{i5} where we show a representation of the shape for the density distribution for the LN (top panel) and the GRPG (central panel). For the LN, we can see that there is a large concentration of intersections in the centre, while the suburbs have a more homogeneous shape characterised by high peaks. For the GRPG, the overall shape does not have any large discontinuities. In both the panels, we notice how the power law effect of the edge length distribution of Eq.\ref{1} produces local inhomogeneous patterns as isolated peaks. This effect is more evident for the LN. The reason is that London grew to incorporate pre-existing town centres. In the bottom panel of Fig.\ref{i5}, we show the contour plot for the intersection density of the LN with the position of the town centres superimposed on this, noting how the density pattern is correlated with them.

In the right panel of Fig.\ref{i6}, we show the comparison of the average length of the road fragments $l(r)$ as a function of the distance from the centre. In this case, we see that the model agrees very well with the real network for the first 15Km. The average increase in the lengths of the edges of the considered graphs is a clear evidence of the growth of both the systems in which on average, the centres of the graphs are filled with short edges and the periphery is sparser where there is space for longer edges. The large fluctuations that are evident in the GRPG model for large values of $r$ are due to finite size effects.

\begin{figure}[!ht]\center
                \includegraphics[width=0.5\textwidth]{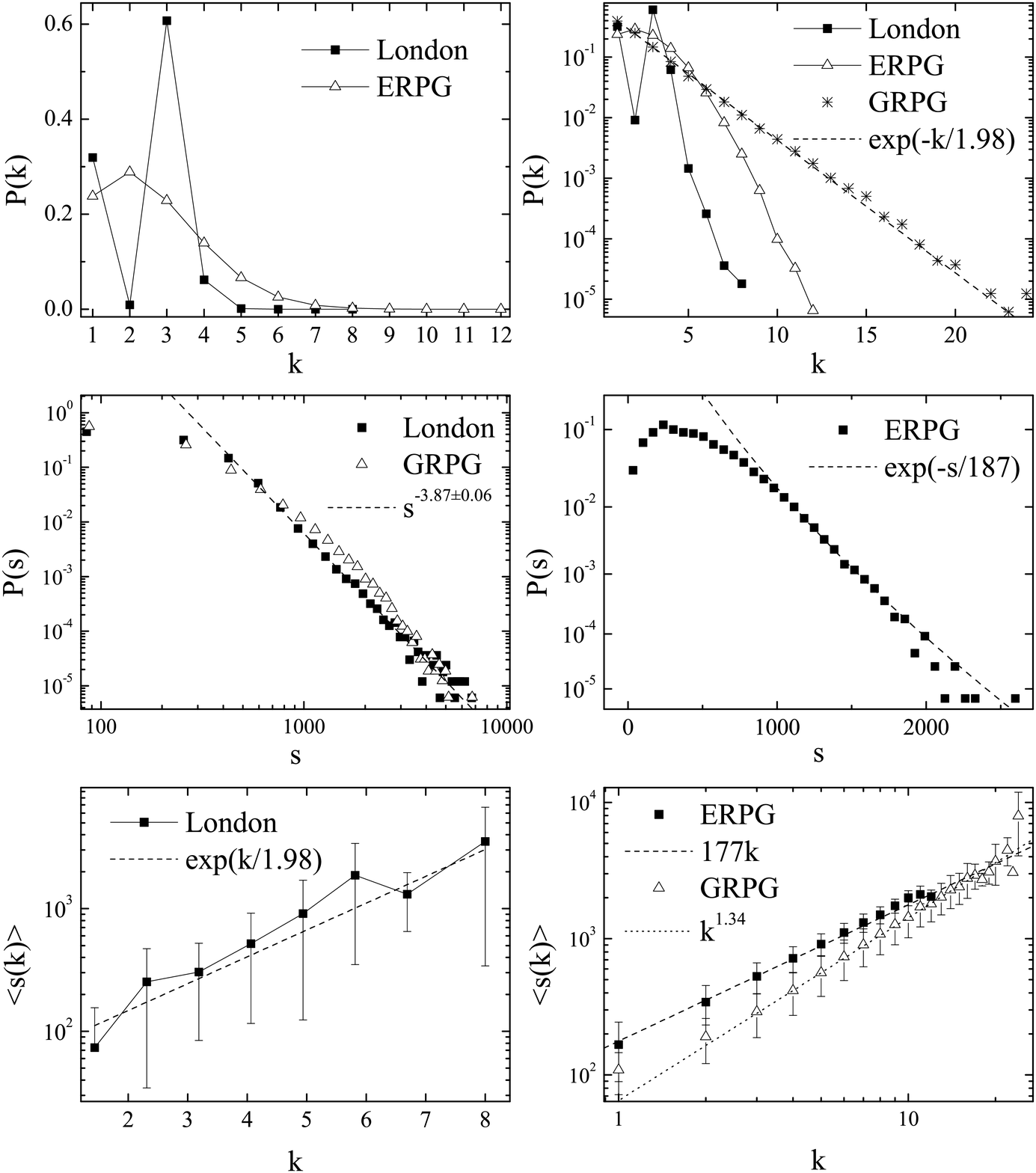}
 \caption{\label{i9}Top left panel: the degree distribution $P(k)$ for the LN and the ERPG. Top right panel: the degree distribution $P(k)$ for the LN, the ERPG  and the GRPG  on a semi-logarithmic scale. The parameter of the exponential function fitting the distribution for the GRPG  has a standard deviation $\sigma=0.02$. Central  left panel: the strength distribution $P(s)$ for the LN and the GRPG  on a double-logarithmic scale. Central right panel: the strength distribution $P(s)$ for the ERPG  on a semi-logarithmic scale. Bottom left panel:  the average strength $<s(k)>$ as a function of the degree $k$ for the LN on a semi-logarithmic scale. Bottom right panel: the same function measured  for the ERPG  and the GRPG  on a double-logarithmic scale. }
    \end{figure}

Even if planar graphs in nature are not characterised by a high degree of connectivity for the vertices, the degree distribution of different planar graphs show non trivial characterisations. For topological aspects, our networks are completely specified by their weighted adjacency matrix $W=\{w_{ij}\}$, where $w_{ij}=l_{ij}$, for $0<i,j\leq V$,  $l_{ij}$ being the length of the street segment connecting vertex $i$ and vertex $j$, if vertex $i$ and vertex $j$ are connected, and $wij = 0$ otherwise. The degree $k_i$ of  vertex $i$ is defined as the number of connections of vertex $i$, $k_i=\sum_{j}\Theta(w_{ij})$ and in this case, it represents the number of streets intersecting at the given intersection. In the top left panel of Fig.\ref{i9}, we show the degree distribution for the LN and the ERPG model using a linear scale. It is worth noting that vertices with degree two were suppressed in the construction of LN. We observe two peaked distributions with a maximum around the average degree, where it is possible to appreciate that the peak for the LN is much higher than the one for the ERPG model. Moreover the maximum degree for the LN is 8 while it is 12 for the ERPG. In the right panel of the same figure, we observe the behaviour of the tail for the same distributions. It seems that they are both ill-defined distributions, very similar to the ones found for ant galleries in \cite{11}, but that to claim they show exponential behaviour would be misleading. In the top right panel of the same figure, we show the degree distribution for the GRPG model. In this case, the distribution is not peaked and the exponential behaviour is clearly distinguished with a maximum degree $k_{max}=24$. This observation is very important. In fact, LN is a growing system and the fact that it does not display an exponential degree distribution relates to its particular organisation more than to its similarities to the ERPG.

In weighted graphs, the strength of vertices often provides important information about the system and is strictly correlated to the degree of the vertices \cite{29}. In our case, the strength $s_i$ of  vertex $i$ is defined as the sum of the lengths of the street fragments intersecting that vertex, $s_i=\sum_jw_{ij}$. In our three samples, the strength measures and their correlations are quite diverse. In the central panels of Fig.\ref{i9}, we show the strength distribution for the LN, the ERPG and the GRPG. For LN (in the central left panel), the strength distribution shows a clear scale-free behaviour with exponent $-3.87 \pm 0.06$. We find a similar behaviour in the GRPG (in the central left panel) even if its scale free behaviour is not well defined, while for the ERPG model (in the central right panel), the strength distribution is a peaked function with an exponential tail.

To understand the correlations between strength and degree of a vertex, in Fig.\ref{i9} we plot the average strength $<s(k) >$ which is measured as a function of $k$. In the case of LN (in the bottom left panel), $<s(k) >$ displays growing behaviour that can be fitted with an exponential curve within the error bars. In the bottom right panel on a double logarithmic scale, we can observe how $<s(k) >$ displays linear growth, $<s(k) >=<l>k$, for the ERPG, where $<l>$ is the average length of the edges . For the GRPG, this shows super-linear growth, $<s(k)>\propto k^{1.34}$,  as observed in many other topological growing networks \cite{30}.

\begin{figure}[!ht]\center

                \includegraphics[width=0.5\textwidth]{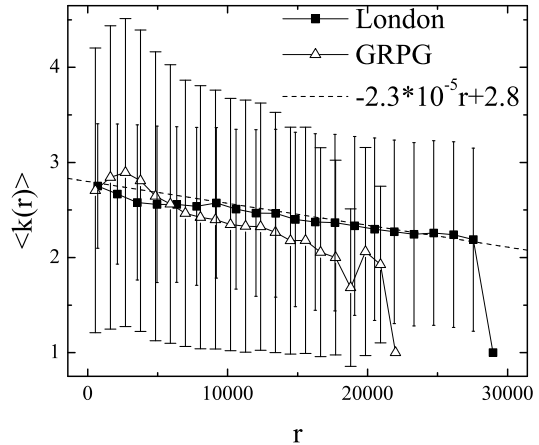}
 \caption{\label{i10} The average degree $<k(r)>$ as a function of the distance from the centre for  LN  and the GRPG. The LN data are well fitted by a linear function. }
    \end{figure}

 The last measure we show in this section is the average degree of the vertices as a function of the distance from the centre $<k(r) >$. This allows us to see how much the topological and metrical spaces are related. In Fig.\ref{i10}, we show $<k(r) >$ for LN and GRPG. In the case of ERPG and GM, $<k(r) >$ is just a constant function of $r$. In the case of LN, $<k(r) >$ decays linearly from the centre to the periphery. In the GRPG, $<k(r) >$ decays more rapidly. This decay function is a signature of the growth of the system where the centre is more densely connected.

\subsection{Measures in the cycle space}

It is interesting to observe a planar graph in its cycle space, that is the space formed by all the edges of the graph that are part of a closed polygon \cite{4}. In fact it is in that space that many of the planar graph properties are best understood.

\begin{figure}[!ht]\center
                \includegraphics[width=0.5\textwidth]{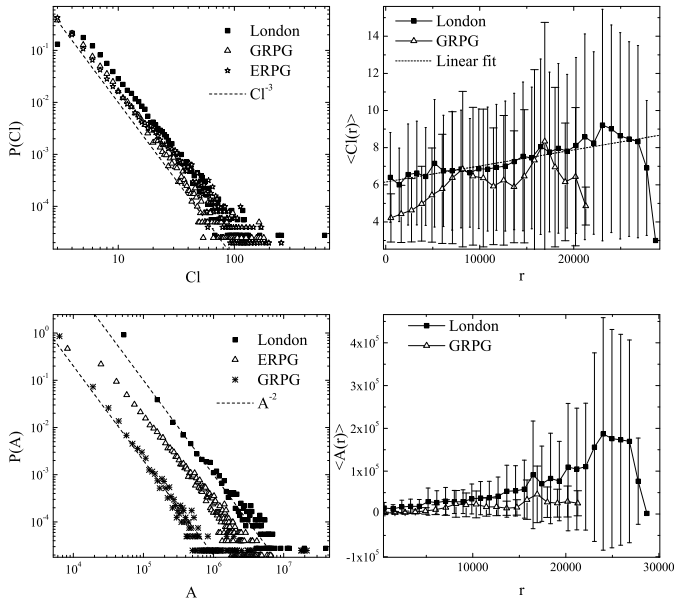}
 \caption{\label{i11} Top left panel: the frequency distribution $P(Cl)$ for the length of the cycles $Cl$ for the LN, the GRPG and the ERPG. The dashed line is a power law with slope $-3$. Top right panel: the measure of the average length of the cycles $<Cl(r)>$ as a function of the distance $r$ from the centre. Bottom left panel: the frequency distribution $P(A)$ for the area of the faces $A$ for the LN, the GRPG and the ERPG. The dashed line is a power law with slope $-2$. Bottom right panel: the measure of the average area of the faces $<A(r)>$ as a function of the distance $r$ from the centre. }
    \end{figure}

The length of a cycle $Cl$ is defined as the number of its edges or vertices which, is an important number in understanding the geometry of the graph. In the GM, the cycle space is trivial. In the top panels of Fig.\ref{i11}, we show the measures related to the cycle lengths in our networks. The top left panel shows the frequency distribution $P (Cl)$ for the cycle lengths for LN, the ERPG and the GRPG. It is interesting to note that this distribution has a power law tail with a very similar slope for three of the networks with exponent -3. The significant differences between LN and the random graphs is that in LN, cycles of length 4 and 5 are more numerous than cycles of  length 3 and that the tail for the LN is much longer than the tails of the random networks. This is probably due to the existence of geographical constraints in that LN growth forces the creation of very large polygons (for instance around the Thames seen from the right panel of Fig.\ref{i2}). In the right panel of Fig.\ref{i11}, we show the measure of the average cycle length $< Cl(r) >$ as a function of its distance $r$ from the centre for both the LN and the GRPG. The ERPG model is not included in the figure since in that case $< Cl(r) >$ is a constant function of $r$. In both the LN and the GRPG model, $< Cl(r) >$ is a growing function of $r$ which is characteristic of growing systems where central polygons are smaller and the average connectivity is larger. Nevertheless the growth of $< Cl(r) >$ for LN is more steady and it is well fitted by a linear function. The decay behaviour of $< Cl(r) >$ for large values of r is due to finite size effects.

The area of the faces $A$ is also a measure used to characterise urban networks \cite{17}. In the bottom left panel of Fig.\ref{i11}, we show the frequency distribution for $P (A)$ for the area $A $ of the faces of LN, the ERPG and the GRPG. For the LN, we find a good agreement with the power law slope measured for the road network of Dresden \cite{17}. Interestingly we also find that other stochastic networks show a similar behaviour to those of London and Dresden, suggesting that the power law behaviour for the faces area distribution is not likely to be a sign of complex self-organisation of an urban system, nor of its growth.

In the bottom right panel of Fig.\ref{i11}, we show the measure of the average area of the faces $<A(r) >$ versus the distance from the centre $r$ for the LN and the GRPG models. For the ERPG, $<A(r) >$ is a constant function of $r$. As we expect, for LN and GRPG, the area of the faces is a growing function of $r$, supporting the hypothesis of a strong mono-centric component in the growth of the city. It is also interesting to note how the fluctuations grow with distance from the centre.

\subsection{Centrality measures}

\begin{figure}[!ht]\center
                \includegraphics[width=0.5\textwidth]{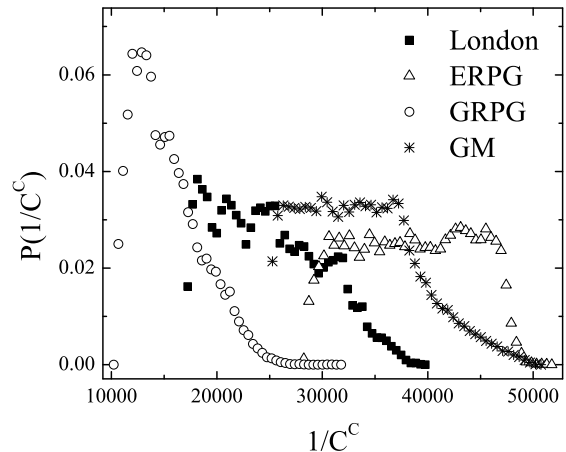}
 \caption{\label{i13} Probability distribution for the inverse of the closeness centrality $P(1/C^C)$ for the LN, the ERPG, the GRPG and the GM.}
    \end{figure}

The closeness centrality measures how much a vertex is to the traffic on the network, that is how much of the network is easily reachable from all its different vertices. It is defined as:
\begin{equation}\label{2}
C_i^C=\frac{V-1}{\sum_{i\neq j}d_{ij}},
\end{equation}
where $d_{ij}$ is the sum of the lengths of the street segments forming the shortest path between vertex $i$ and vertex $j$. We believe that the inverse of Eq.\ref{2}, $1/C_i^C$, measured in Km, gives a better understanding of the dynamics of those   networks, since it represents the average metric distance between the intersection $i$ and all the other intersections of the graph . This gives an effective understanding of the physical effort (the informational effort will be considered in the next section) expended in navigating a city. In Fig.\ref{i13}, we show the distribution $P(1/C^C)$ for each of our networks. We can also see how the networks are highly differentiated by this measure. The ERPG is the one which is less \emph{travel friendly}, the vertices being more distant on average from all the other vertices, even if we only consider the connected portion. The majority of vertices lie on a plateau between 30Km and 46Km, that are the values where these vertices are uniformly distributed.

On a travel friendly scale, the ERPG has lower centrality than the GM. The distribution is similar, presenting a large plateau, but the plateau for the GM is now higher and thinner, between 26Km and 37Km and the tail falls exponentially for more than 10 Km. Still considering the travel friendly scale, the centrality of LN lies between the static and growing models. For the LN,  a plateau does not really exist. After peaking around 18Km, the distribution decreases with many fluctuations, but with an overall linear trend, to a maximum average distance of 40km. Then it appears that the most travel friendly pattern is that given by the GRPG, where we find a large peak around 13 km and a fast and smooth decay until 32km. The extension of the GRPG city is smaller than in the other models, as we have already noted. What is interesting is the lack of a plateau for both the LN and the GRPG.

\section{The dual representation and the alignment problem}

The network of urban streets shows scale-free properties for its degree distribution when it is  considered in its dual representation, where the vertices are the streets and two vertices are connected if the streets they represent intersect \cite{1}. This is important for it allows us to look at the growth of cities in a novel way through an informative perspective. In this section, we examine in detail the properties of the dual street network of London (hereafter DLN) and we compare it to the properties of the dual representations of the three other models that we have introduced as our idealised baseline.

\begin{figure}[!htbp]\center
         \includegraphics[width=0.5\textwidth]{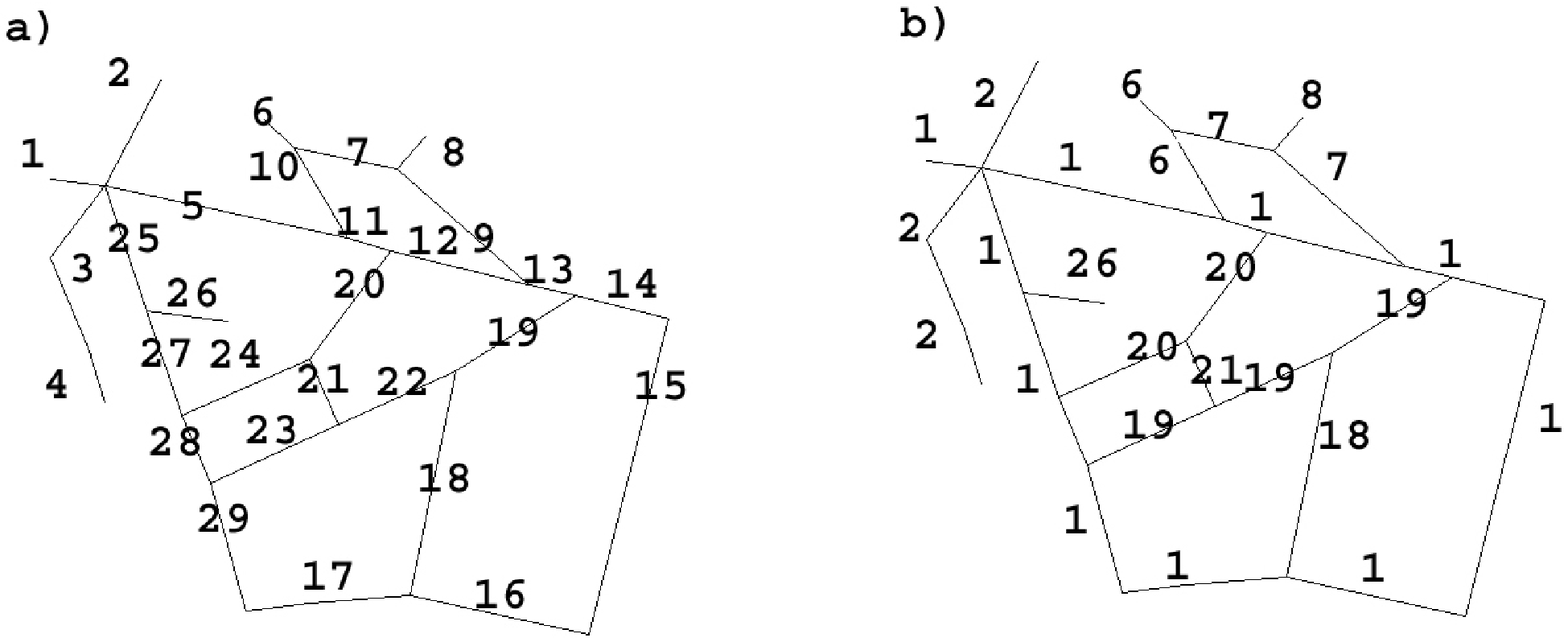}
         \includegraphics[width=0.26\textwidth]{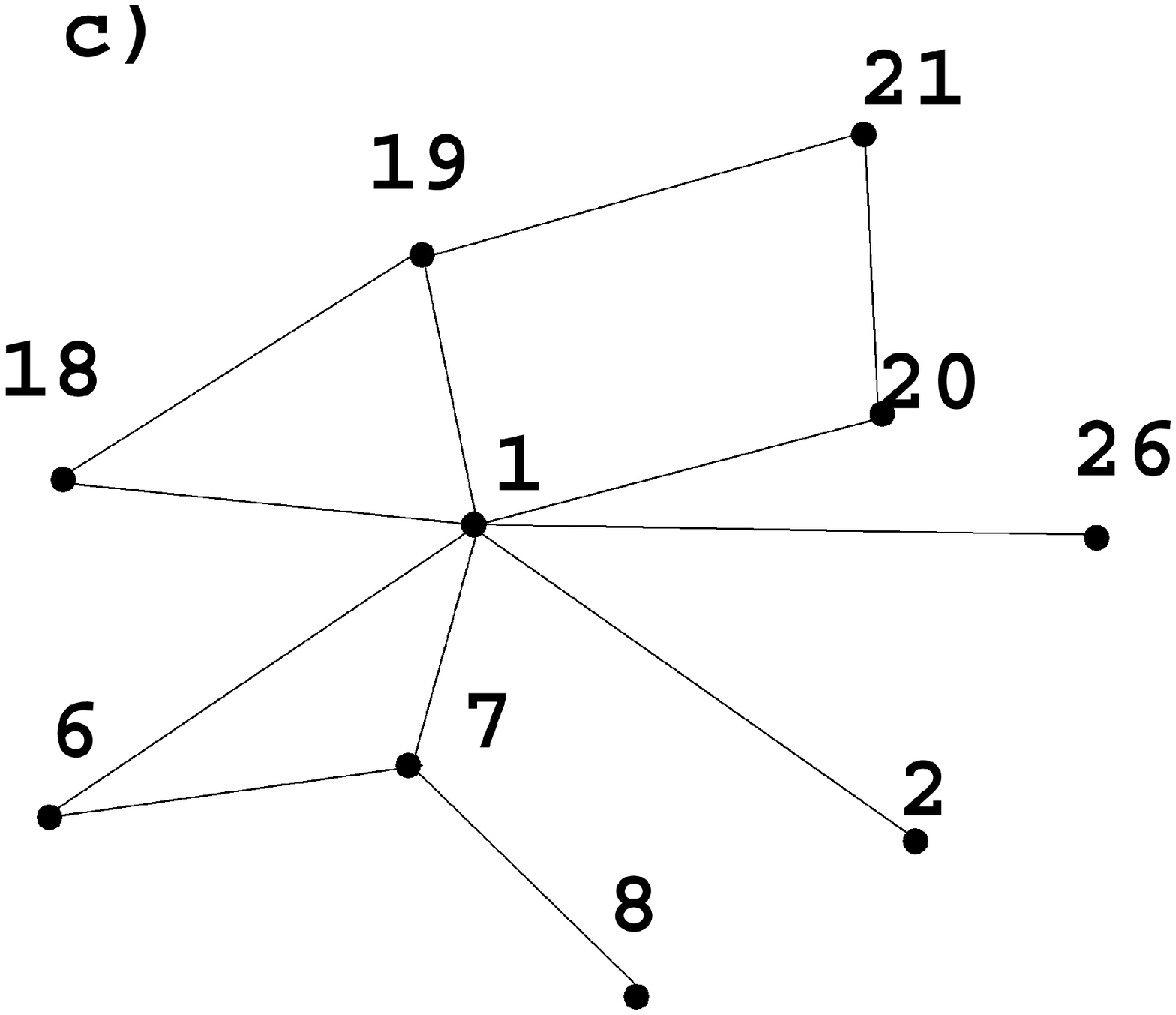}
          \caption{\label{f2} The process to create a dual graph from a street graph. Panel \emph{a}: a fragment of the street graph of London, every different street segment has a different ID. Panel \emph{b}: the same street network where the street segment ID are changed after applying an alignment principle (the ICNP). Panel \emph{c}:  the final dual graph representation of the street network of Panel \emph{b}, where the street IDs become the vertices and two vertices are connected if the streets they represent intersect. }
\end{figure}

The procedure to build a dual street network is to assign the same label or ID to the street fragments that belong to the same road using an alignment principle. Then the dual representation of a planar graph is a network in which the roads are vertices and two vertices are connected if the roads they represent intersect. The procedure used to obtain a dual graph from a planar graph is shown in Fig.\ref{f2}. In that construction, long roads with the same ID connect to many roads, while short roads such  as dead-ends, connect to just one or a very small number of other roads. In this way, hubs form at all scales producing a characteristic shape of the degree distribution that is common to many self-organising systems \cite{15}.

An important issue is to find an algorithm to establish how different street fragments might belong to the same street, i.e. an algorithm to assign the ID to the different street fragments. In \cite{2}, a \emph{name-street approach} is considered where two street segments are given the same ID if they have the same street name. Unfortunately, as noted in \cite{8}, this approach does not consider the fact that in many cities, many streets share the same street name without intersecting. Also it is possible to find the same physical streets that have two or more separate names. London is rich in both of these phenomena. In our view, the efficient approach called \emph{Intersection Continuity Negotiation} (ICN) is worth considering \cite{3}. This approach starts from the principle that two street fragments belong to the same road if the angle they form is close to 180 degrees. Then the procedure of the ICN is to rank pairs of street fragments at a given intersection by the convex angle they form. Then the same ID is given to the street fragments that form the major convex angle.

\begin{figure}[!htbp]\center
         \includegraphics[width=0.4\textwidth]{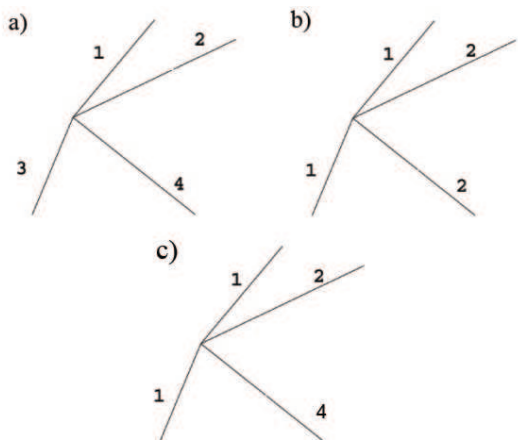}
          \caption{\label{f1}  Panel \emph{a}: a generic crossroad with random labels. Panel \emph{b}: the same crossroad where IDs are reassigned by ICN principle. Panel \emph{c}: the same crossroad where IDs are reassigned by ICNP principle.}
\end{figure}

This approach is very efficient but in our view, it fails to correctly describe the situation shown in Fig.\ref{f1}.
Fig.\ref{f1} shows how the ICN principle assigns the ID to different streets at an unusual crossroad. Referring to the figure, it seems more plausible to use a negotiation principle that transforms street segment 3 into 1 as ICN does leaving the other ID unchanged (panel \emph{c} of Fig.\ref{f1}). This situation appears in reality when dealing with a ring road or a beltway where other roads enter or exit.

To fix this problem, we have extended the ICN principle to the ICN Plus (ICNP) principle in which the ICN is considered in all the cases in which the larger convex angle is formed between road segments that are not adjacent. In the case where two adjacent road segments form the largest convex angle as in Fig.\ref{f1}, we give them the same ID. But we do not change the ID of the other street segments intersecting the vertex, considering them as different roads. A more precise description of the ICNP algorithm is given in appendix B.

The networks obtained in this way are unweighted and undirected. They are called \emph{information networks} since they describe the way people think about moving in a city. To go from one point to another in a city, we do not need to know all the street segments and intersections that join the two points but only the name of the roads that enable us to navigate. For instance, considering the top panels in Fig.\ref{f2}, if we want to travel from street segment 1 on the extreme left to the street segment 19, we would normally go straight along road 1 and then turn right onto road 19 as shown in panel \emph{b} and not go straight along line 1, then taking line 5, line 11, line 12, line 13 and eventually turning right onto line 19 as shown in panel \emph{a}. So we can say that to go from line 1 to line 19, just one unit of information is required, as is clear from panel \emph{c} of the same figure. Then the maximum information required to cross a city is the diameter of its information network, not the diameter of its primal network, where the diameter $D$ of a network is defined as the maximal shortest path connecting two vertices of that network.

It is worth noting that whatever algorithm is used to create the dual representation, it always contains bias. In our case, the longest road recognised has a length of around 17Km. The orbital M25, for example, is not recognised as a single road, nor are other important routes such as the A40, connecting the centre of London to Oxford. These biases are then reflected in the degree distribution whose exponential tail is not well understood. Other solutions to the alignment problem are possible and research on the topic is active \cite{31}.

\section{Dual analysis}

In this last section, we will draw the discussion to conclusion by  analysing the properties of the dual representation of LN, of the ERPG (hereafter DERPG), of the GRPG (hereafter DGRPG) and of the GM (hereafter DGM). These networks are purely topological, in the sense they are not embedded in Euclidean space per se. We will split the section in two parts. In the first part we examine the main topological properties of those binary networks such as degree distribution, clustering coefficient and nearest neighbour degrees, and in the second part, we analyse the network using an informational approach through centrality measures.

\subsection{Topological properties}

\begin{figure}[!htbp]\center
         \includegraphics[width=0.5\textwidth]{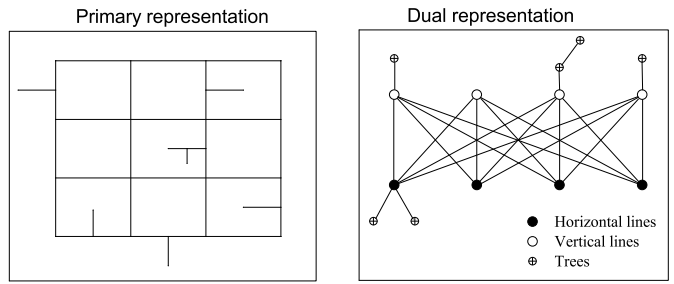}
          \caption{\label{f7} The dual representation of the Grid Model (left panel) is a bipartite graph (right panel) in which horizontal lines and vertical lines become different families of vertices. }
\end{figure}

In Tab.\ref{t1}, we present the main topological properties for the dual representation of the considered networks. In the table, we show the number of vertices $V$, the number of edges $E$, the average degree $<k>$, the diameter $D$ and the average clustering coefficient $<C>$ for the four networks. We can already observe how different topologies in the primary representation give rise to very different dual networks. Remembering that in the dual representation, the number of vertices is the number of different roads and that the number of edges is the number of intersections between different roads, we see that in the DLN there are a larger number of roads than those generated in the random networks. In spite of this, the diameter of the DLN is much smaller than the diameter of the random networks, a diameter that has the size of the same order of the logarithm of the number of vertices which is a small world property \cite{24}. This means that even if random roads are longer than real ones, they are not organised to fill the space as efficiently as in the DLN. This effect is very much related to the angular distribution of the edges at the intersections of the primary graphs that generate the big differences in the average clustering coefficient to be explained below. In the case of the DGM, it has already been shown  \cite{8}  that this is a bipartite graph in which one family of vertices represent the horizontal lines and the other family represent the vertical lines. Every vertex of one family is connected with all the vertices of the other family. From those vertices, small trees are generated (as in Fig.\ref{f7}) which represent the $m$ lines added to the GM (see the introduction) to obtain the desired average degree.

\begin{table}[htbp]
\begin{center}
\begin{tabular}{|l|ccccc|}
\multicolumn{6}{c}{Dual representation}\\\hline
&V&E&$<k>$&D&$<C(k)>$\\\hline
\emph{DLN}&74782&107988&2.89&33&0.042\\
\emph{DERPG}&54458&91732&3.36&243&0.31\\
\emph{DGRPG}&67052&222374&6.73&72&0.44\\
\emph{DGM}&64296&100250&3.12&13&0\\\hline
\end{tabular}
\end{center}
\caption{Number of vertices $V$, number of edges $E$, average degree $<k>$, diameter $D$, and average clustering coefficient $<C(k)>$ for the DLN, the DERPG, the DGRPG and the DGM.}
\label{t1}
\end{table}

\begin{figure}[!htbp]\center
         \includegraphics[width=0.5\textwidth]{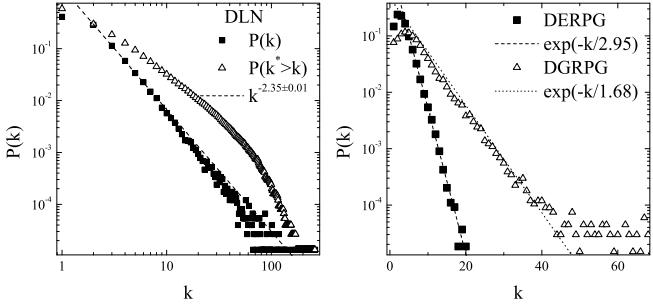}
          \caption{\label{f3} Left panel: degree distribution $P(k)$ and cumulative degree  distribution $P(k^*>k)$ for the DLN on a double-log scale. Right panel: degree distribution for the dual network of the DERPG model and the dual network of the DGRPG model on a semi-log scale. The fat tail of the latter has been cut in this plot, but it appears at $k_{Max}=229$. }
\end{figure}

In the left panel of Fig.\ref{f3}, we show the degree distribution $P (k)$ and the cumulative degree distribution $P (k* >k)$ for the dual network of London. The exponent of the best fitting line has been calculated for the cumulative distribution. From the degree distribution, we can see that a power law behaviour emerges with a fat tail. From the cumulative distribution, we can see how the tail of this distribution falls faster for large values of the degree. The same behaviour has been observed at national scales \cite{18} which can been attributed to the natural boundaries of the UK viewed as an island. We can say the same thing in this case where a natural cut-off emerges for the finite sample size. For the tail, we also have to consider the above mentioned biases due to the choice of the alignment principle in the construction of the dual graph.

In the right panel of Fig.\ref{f3}, we show the degree distribution $P (k)$ for the DERPG and the DGRPG on a semi-log scale. Notably the maximum degree for the DERPG is $kmax = 20$ compared to $kMax = 261$ for the DLN and $kMax = 229$ for the DGRPG. We argue that the static planar graph has a structure that does not allow long ``roads" to form, while the tree growing structure of the GRPG gives rise to ``roads" with a length comparable to the ones of the LN. Moreover it is interesting to note how the distribution of the random networks is radically different from the LN in terms of its information space. In the stochastic models, we observe an exponential behaviour for the degree distribution. In the case of the DGRPG, this exponential behaviour encapsulates a fat tail that appears at a maximum degree $kMax = 229$. We are tempted to speculate that this exponential behaviour relates to the lack of informational organisation of such random systems.

\begin{figure}[!htbp]\center
         \includegraphics[width=0.5\textwidth]{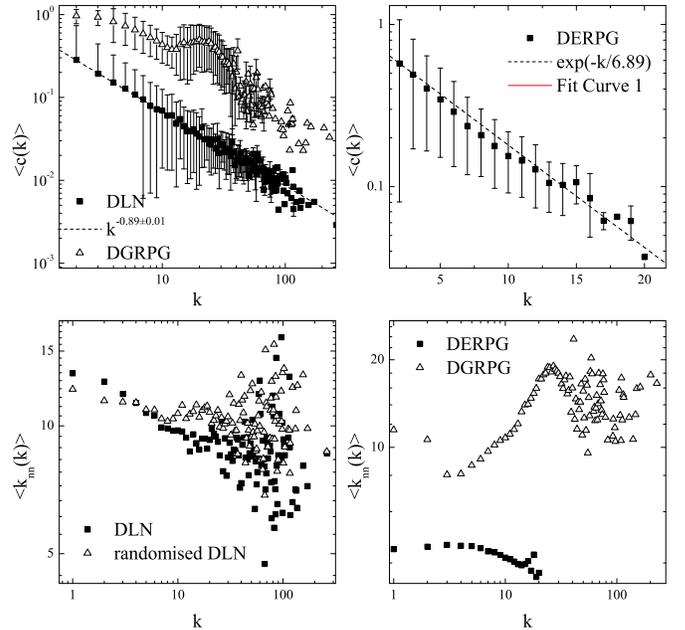}
          \caption{\label{f4} Top panels: average clustering coefficient $<c(k)>$ versus the degree $k$ measured in the DLN and the DGRPG  (left panel) and for the DERPG  (right panel). Bottom panels: average nearest neighbours degree as a function of the degree $<k_{nn}(k)>$ measured in the DLN and the randomised DLN (left panel), the DERPG  and the DGRPG (right panel). }
\end{figure}

The \emph{clustering coefficient} or \emph{transitivity} $c_i$ for a vertex $i$  is the ratio between the number of edges $e_i$ connecting each other the nearest neighbours of vertex $i$ and the number of such possible edges, and it is defined as:
\begin{equation}
c_i(k_i)=\frac{2e_i}{k_i(k_i-1)}.
\end{equation}
The average clustering coefficient $<c(k) >$ then counts the number of triangles in the graph. In the top panels of Fig.\ref{f4}, we show the average clustering coefficient $<c(k) >$ as a function of the degree $k$ measured in our networks. In the left panel, measures for the DLN and the DGRPG are shown. The average clustering coefficient for the DLN follows a power law with exponent $-0.89 \pm 0.01$. This  behaviour has already been noted in \cite{3} for most of the 1 mile-square samples considered. This effect at a larger scale makes it a characteristic signature of the dual representation of an urban network. This scaling behaviour is well explained by the very low average clustering coefficient $<c>\approx0.04$. This means that in the dual representations only a few triangles form and the larger the degree of a node, the less is the probability that its neighbours are interconnected. The poor triangular structure of the dual representation of urban street network reflects the angular structure of the primary graph where roads tend to be orthogonal and where cycles of length 4 or 5 are more likely to happen than cycles of length 3 (see top left panel of Fig.\ref{i11}).
In the same panel, the average clustering coefficient $<c(k) >$ as a function of k is shown for the DGRPG. The values for $<c(k) >$ are much larger than the ones found in the DLN, with an average clustering coefficient $<c>\approx0.4$, an order of magnitude larger than that for the DLN. That is due to the greater probability in the random network for triangles to form (noting that triangles in the primary space correspond to triangles in the dual space). The behaviour of $<c(k) >$ is now less smooth and not well defined and its tail is much steeper than that in the DLN.

 In the top right panel of Fig.\ref{f4}, we show the average clustering coefficient $<c(k) >$ as a function of $k$ for the DERPG. The average clustering coefficient is $<c>\approx0.31$, still an order of magnitude larger than the DLN, confirming the fact that there are more triangles in a random planar network than in an urban planar graph. Interestingly the shape of the measured function decays exponentially with the degree.

The \emph{average nearest neighbours degree} as a function of the degree $<k_{nn}(k)>$ quantifies the second order correlations of complex networks and is defined as:
\begin{equation}
<k_{nn}(k_i)>=\sum_{k_j}k_jP(k_j|k_i)
\end{equation}
where $P(k_j|k_i)$ is the conditional probability that a vertex with degree $k_i$ has a neighbour with degree $k_j$.
 We have to be careful to analyse the measures of $<knn(k) >$. In fact it has been shown that such networks reveal structural correlations that are due to the degree distribution and to its cut off for large degrees \cite{27}. Hence, for understanding the correlations of the system, it is important to compare the actual $<knn(k) >$ with the one obtained in a randomised network. In the bottom left panel of Fig.15, we show the measure of $<knn(k) >$ as a function of the degree k for the
DLN and the same measure for a network derived from DLN by rewiring all the edges and keeping the degree sequence unchanged. In this way we can see that in the DLN, there are disassortative correlations for small values of the degree, where small degree vertices tend to be connected with high degree ones, while for larger degrees, the network looks uncorrelated. In the bottom right panel of Fig.\ref{f4}, we show the same measure for the ERPG and the GRPG. The former shows a structural disassortative behaviour while the latter shows a structural assortative behaviour, where high degree vertices tend to connect to high degree vertices.

\subsection{Centrality measures}

\begin{figure}[!htbp]\center
         \includegraphics[width=0.5\textwidth]{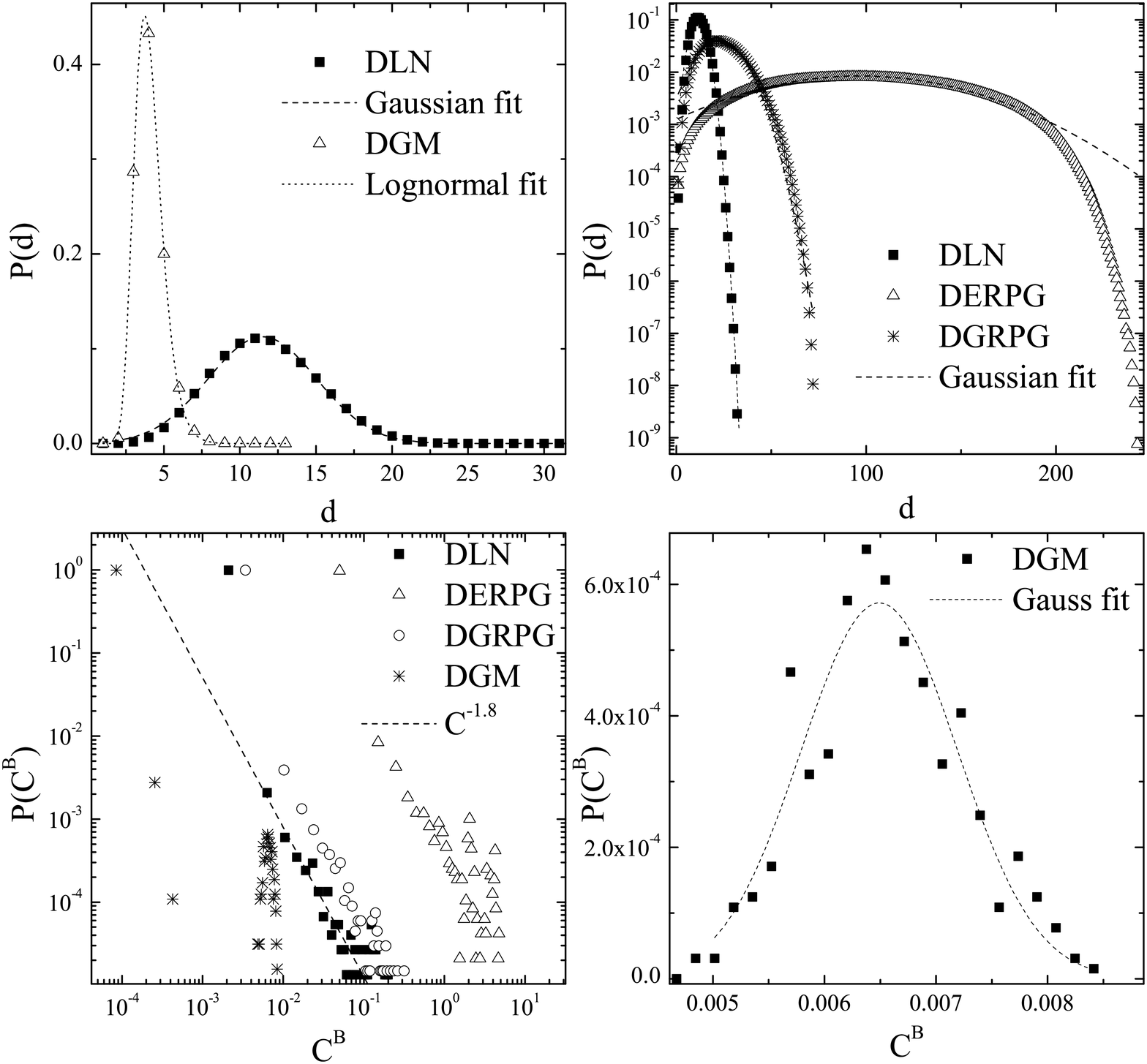}
          \caption{\label{f6} Top panels: distribution $P(d)$ for the average minimal path $d$ between all the pairs of vertices of the network. Left panel: comparison between DLN and the DGM. The DLN distribution is well fitted by a Gaussian distribution (reduced $\chi^2=6.8*10^{-6}$
), while the DGM by a lognormal distribution (reduced $\chi^2=6.1*10^{-6}$). Right panel: comparison between the DLN, the DERPG and the DGRPG. A semi-log scale is applied to better resolve the tails and Gaussian fits are performed to clarify the deviations. Bottom panels: in the left panel the distribution $P(C^B)$ for the betweenness centrality $C^B$ for the DLN, the DERPG, the DGRPG and the DGM. In the right panel a particular view of the measured $P(C^B)$ for the DGM fitted by a Gaussian distribution.}
\end{figure}

The \emph{shortest path} $d_{ij}$ from vertex $i$ to vertex $j$ is defined as the number of edges that form the geodesic that connects vertex $i$ to vertex $j$ and we have that $1\leq d_{ij}\leq D$, where $D$ is the diameter of the graph. In the top panels of Fig.\ref{f6}, we show the frequency distributions $P (d)$ for the shortest paths measured between each pair of vertices in our networks. This is a very important measure since it quantifies the informational content of the network where  $d_{ij}$ represents the mental effort we incur in navigating a city. In this context, we can see how the distribution for the DLN is displaced in between the DGM, the easiest ``city" to navigate, and the random models, the most difficult ``cities" in which to move. In the top left panel of Fig.\ref{f6}, we show $P (d)$ for the DLN and the GM. In the case of the DLN, $P (d$) is well fitted by a Gaussian distribution centred at $p_c=11.74\pm0.05$, with a width or variance of $\sigma=7.14\pm0.09$. $pc$ represents the average information required to move from one point to another in the city.

For the DGM the distribution $P(d)$ is well fitted by a lognormal distribution centred in $pc=3.940\pm0.006$ with width $\sigma=0.230\pm0.001$. That means that the average information to travel in a grid-like city is much less than the one we find IN a large city like London as we could have been expected. In the right panel of the same figure, we show the measures of $P (d)$ for the DLN and the random networks. A semi-log scale is used to better resolve the tails of the distributions. For the random networks, we find that the distributions are shifted to the right in respect of the DLN, meaning that the information required to travel between two random vertices is larger than in the real network.

$P(d)$ for the DGRPG is still well fitted by a Gaussian distribution even if we can see that  the tale behaves slightly differently. The centre of the distribution is $p_c=22.70\pm0.02$ and the width $\sigma=20.1\pm0.3$. The case of the DERPG is interesting too for the measure computed on the connected part of the network is smaller than other networks under consideration. Still the informational content of the network is smaller than those found in the other networks, in the sense that to navigate the ERPG, much more mental effort is needed. We can see from the figure how the tail of the distribution decays faster than the Gaussian curve. The centre of the distribution is at $p_c=96.9\pm0.3$ and its width $\sigma=99.0\pm0.7$. We believe that the reason why the GRPG has more informational content than the ERPG is that the GRPG grows as a tree and this growth gives additional information content for navigation of the network.

The \emph{betweenness centrality} $C^B_v$ for vertex $v$ is defined as
\begin{equation}\label{bc}
C^B_v=\frac{2}{(V-1)(V-2)}\sum_{i<j,i\neq v\neq j}\frac{g_{ivj}}{g_{ij}}
\end{equation}
where $g_{ij}$ is the number of geodesics (shortest paths) connecting vertices $i$ and $j$ and $g_{ivj}$ is the number of geodesics connecting vertices $i$ and $j$ that contain vertex $v$. $C^B_v$ is 0 if $v$ has degree one, that is if it represents a dead end road.  The normalisation factor takes account of the fact that the maximum value for the betweenness centrality is achieved for the central vertex of a star graph \cite{25}. Hence $C^B_v$ is a measure of how probable it is to travel on a certain road when moving from  one a road to another in a city. The distribution function $P(C^B)$ describes the hierarchy of betweenness centrality, if any exists. In the bottom left panel of Fig.\ref{f6}, we show $P(C^B)$ measured for our dual networks. Again we see that this measure provides a good classification for the different networks. In particular, we see that in the DLN, in the DERPG and in the DGRPG a scaling distribution emerges, implying a hierarchy in the centrality of the roads. For the DGM, we observe a scaling relation for low values of CB related to the tree structures of the DGM (see Fig.\ref{f2}). Then $C^B$ is Gaussian distributed around a well defined average (see bottom right panel of Fig.\ref{f6}) and this implies that in a grid, the information content of roads is nearly equivalent. The values assumed by $C^B$ are related to the number of different equivalent geodesics $g_{ij}$ that join different roads,  where as $g_{ij}$ increases, $C^B$ falls. In this sense, we can understand the displacement of the distributions, the extremes being the DGM, where many equivalent geodesics exist between
two roads, and the DERPG where not many different choices exist in traveling from one point to another in the graph. In the between, we find that the DLN and the DGRPG have similar behaviour. We thus believe that the tree growing structure of the DGRPG is very important in reproducing the hierarchy of the betweenness centrality associated with  roads in the DLN.

\section{Conclusions}

A network theory approach to the study of planar graphs and urban networks is a natural consequence of the study of growing cities that fill their space in the manner of self-organising systems but it has not been widely explored to date. Indeed in this paper, we are the first to demonstrate how this can be useful for providing a description of urban growth. Many of the concepts that are crucial in urban planning, such as accessibility and density used in measuring urban sprawl find natural definitions in the interplay between these primary and dual representations of urban systems \cite{32}.

In this paper, we have begun a deeper analysis of street networks for large cities where we develop both primal and dual representations. To contextualise these results, we considered three models for generating planar graphs based a grid, a static planar graph and a growing planar graph. To our knowledge, this is the first time that a growing planar graph has been introduced for this kind of urban analysis, where we have illustrated that many geometrical and topological features of the LN are emerging properties of a growing system and that the GRPG is the best null model for understanding correlations and properties of the LN.

In its primary representation, we found that the degree distribution of the LN is not a trivial outcome of the planarity criteria. It is quite different from the exponential degree distribution that we found for the GRPG, and this is clearly a result of more complex underlying organisation principles. Also in its primary representation, we have explored its topological and geometrical properties and these measures in the cycle space contra that the properties of planar graphs provide a richer texture for description and analysis than envisaged hitherto.

In its dual representation, we have had the opportunity to observe how a real system is very different from a random one in terms of its information space. Interestingly we found that if the degree distribution of the DLN is scale free, those in random planar graphs are exponential. This means that the scale free distribution found in urban structure is a signature of a complex organisation within the information space and the parallels with classical topological networks behaviour are thus straightforward \cite{15}.

In our view, the underlying principles of the organisation of the street network of large cities like London can be framed through a comparison of the centrality measures in their primary and dual representations. In fact, we found that while the GM is easy to navigate in terms of its information space, it is costly to navigate in metrical space, and while the GRPG is easier to navigate in its metrical space, it is difficult to navigate in its information space. Thus the LN appears to be a self-organised compromise between those two models, a system that balances the effort in spatial displacement which attempts to minimise the amount of information that acts to generate that displacement. Further developments of this research will pursue the applicability of this network model in developing descriptions and analysis of urban systems that reflect least effort principles. Moreover a more detailed analysis of growing random planar graphs with different arc length distributions will be interesting to understand more general properties of growing random planar graphs.

\section{Acknowledgments}

This research was part funded by the EPSRC SECSE project (www.secse.net/) and by the GLA (Greater London Authority) Economics Division. We wish to thank Konstantin Klemm for the useful discussions about the cycle space. Many of the measures in this paper were done using the freeware $R$ package \emph{IGraph} developed by G\'{a}bor Cs\'{a}rdi and Tam\'{a}s Nepusz \cite{34}.

\begin{appendix}

\section{The London street network}

The London street network was derived from two Ordnance Survey (OS) dataset products \cite{33}, OS MeridianTM 2 which includes Motorways, A Roads, B Roads and Minor Roads, and the OS Integrated Transport Network (ITN). The latter includes all the above roads but in more detail with respect to a much greater number of minor roads. The reason two networks were used was that the ITN layer contains more detailed street geometry such as traffic islands and roundabouts and therefore more edges and vertices. Many of these were not needed for the analysis as we were only interested in roads connecting to other roads but this data provided the detail needed for construction of the full network. For example, each lane entering into the roundabout was represented as a separate vertex while traffic islands have two edges and two vertices. To reduce the number of vertices and edges, roads in the ITN layer that were represented in Meridian dataset were removed (through a buffering operation). This left only the minor roads that where not part of Meridian network which could then be snapped to the Meridian network.

\section{The ICNP algorithm}

We start with a planar graph $\mathfrak{G}=\{V,E\}$ in which every edge has a different label or ID, represented by an integer number $1\leq ID\leq E$. We randomly pick an edge of the graph and for each of its vertices, we consider all the edges intersecting at the vertex. Then we rank pairs of edges according the maximum convex angle they form. We next consider the pair of edges forming the larger convex angle and we relabel the edge with major ID giving to it the ID of the other edge. We repeat this operation for the remaining edges at the intersection. If the number of edges at the intersection is odd, then the last edge in the hierarchy of convex angles is not relabelled. If the edges forming the major convex angle are adjacent, then we relabel them according to the above description, and we leave the ID of all the other edges at the intersection unchanged.  We repeat this process for $N\sim E^{3/2}$ times.

\end{appendix}

%

\end{document}